%% file: main.tex
\documentclass[aps,prb,twocolumn,superscriptaddress,showpacs,floatfix]{revtex4-2}
\usepackage{comment}
\usepackage{physics}
\usepackage{amsmath,amsthm,amssymb,amsfonts, color, comment, graphicx}
\usepackage{xcolor}
\usepackage{float}
\usepackage{graphicx}
\usepackage{dcolumn}
\usepackage{bm}
\UseRawInputEncoding 
\usepackage{dsfont}
\usepackage{hyperref}

\input{defns}

\usepackage{graphicx}
\usepackage[caption=false]{subfig}

\newcommand{\contribute}{$^{\phi}$}

\footnotetext{\contribute~These authors contributed equally to this work}

\begin{document}
\preprint{APS/123-QED}
\title{Jahn-Teller effect in $j\!=\!3/2$ Mott insulators: Ground states and thermal fluctuations}
\author{Kathleen Hart\contribute}
\email{kf.hart@mail.utoronto.ca}
\affiliation{Department of Physics, University of Toronto, 60 St. George Street, Toronto, ON, M5S 1A7 Canada}
\author{Ruairidh Sutcliffe\contribute}
\email{ruairidh.sutcliffe@mail.utoronto.ca}
\affiliation{Department of Physics, University of Toronto, 60 St. George Street, Toronto, ON, M5S 1A7 Canada}
\author{Arun Paramekanti}
\email{arun.paramekanti@utoronto.ca}
\affiliation{Department of Physics, University of Toronto, 60 St. George Street, Toronto, ON, M5S 1A7 Canada}
\date{\today}
\begin{abstract}
The interplay of strong atomic spin-orbit coupling with Jahn-Teller (JT) lattice distortions is an
important theme in quantum materials hosting heavy atoms. A prototypical example of such a system is a degenerate
$j\!=\!3/2$ multiplet coupled to local phonon modes. Here, we study the multipolar ground states of this system
as realized in Mott insulators,
explore its thermal phase diagram via an $SU(4)$ spin Monte Carlo approach coupled to JT phonons,
and study the temperature dependent splittings of the $j\!=\!3/2$ multiplet as relevant to
spectroscopic probes. Our work sheds 
light on coexisting distinct multipolar orders engendered by phonon coupling, role of thermal JT fluctuations,
and the entropy of the coupled multipole-phonon system. We discuss broad implications for double perovskites
Ba$_2$MgReO$_6$, Ba$_2$NaOsO$_6$ and lacunar spinels such as GaTa$_4$Se$_8$ and GaNb$_4$Se$_8$.
\end{abstract}
\maketitle


{\it Introduction.---} 
The Jahn-Teller (JT) effect is a key manifestation of electron-lattice coupling in systems with electronically degenerate states,
wherein a symmetric configuration spontaneously distorts to lower the total energy \cite{jahn1937stability}. 
This symmetry-lowering distortion 
lifts the electronic degeneracy and plays a central role in sculpting the structural, electronic, and magnetic properties of a wide range of 
molecules and correlated solids \cite{kugel1982jahn,jahn1937stability}. Recent work on solids hosting heavy transition metal ions has shown that relativistic
spin-orbit coupling (SOC) can
fundamentally alter electron-lattice interactions, enhancing or quenching JT distortion, or modifying its
potential energy surface, depending on the
electronic filling \cite{Streltsov2020Jahn,Streltsov2022Interplay}. 
Furthermore, strong coupling to JT phonon modes can engender exotic superconductivity \cite{hotta2010weak,gunnarsson1997superconductivity,keller2008jahn},
multiferroic orders \cite{dong2019magnetoelectricity,song2019lattice} and small polaron formation \cite{grimaldi2010large,celiberti2024spin,xing2020localized,franchini2021polarons}. 
In addition, it can play a pivotal role in probing or controlling
spin-orbital entangled degrees of freedom \cite{Maharaj2017_strain,Ikeda2018_strain,patri2019unveiling,patri2020theory,khaliullin2021exchange,ye2023_octupolestrain,ning2023coherent,voleti2023probing,hart2024phonon,yao2025theory,juraschek2022giant,sutcliffe2025n}.

A broad class of quantum  materials highlighting this interplay of SOC with JT phonons
are those hosting $j=3/2$ states which exhibit a
wide range of emergent phases such as Luttinger semimetals \cite{Moon2013_LuttingerSM,Kondo2015_LuttingerSM,Link2020_LuttingerSM},
topological superconductors 
\cite{boettcher2016superconducting,brydon2016pairing,roy2019topological,sim2020multipolarsuperconductivity,ishihara2021tuning}, surface superconductivity and
stripe orders in KTaO$_3$ \cite{Liu2021_KTaO3,Liu2023_KTaO3,villar2021striped,Buessen2021_KTaO3}, and coexisting spin-lattice
orders in lacunar spinels \cite{jeong2017direct,magnaterra2024quasimolecular,Ishikawa2020Nonmagnetic,Yang2024Jahn,Geirhos2021Cooperative} and
double perovskites \cite{Liu2018Nature,Hirai2020Detection,Tehrani2021Untangling,Iwahara2023Vibronic,soh2024spectroscopic,FMosca2024Interplay}. 
Since lattice distortions which lower the local point group symmetry split the degenerate $j\!=\!3/2$
quartet into two doublets, these systems generically exhibit a strong coupling to 
JT phonons.
In addition, the $j\!=\!3/2$ multiplet supports diverse multipolar degrees of freedom
--- dipolar, quadrupolar, and octupolar --- which can lead to hidden
order states with nontrivial topology and higher-order nonlinear responses \cite{you2021multipolar,sorn2024signatures,xie2021higher}.
Thus, understanding the
interplay of SOC and JT effects in this class of materials can broadly shed light on the ground states and thermal
fluctuation effects in strongly coupled multipole-phonon systems.



\begin{figure}
    \centering
    \includegraphics[width=0.98\linewidth]{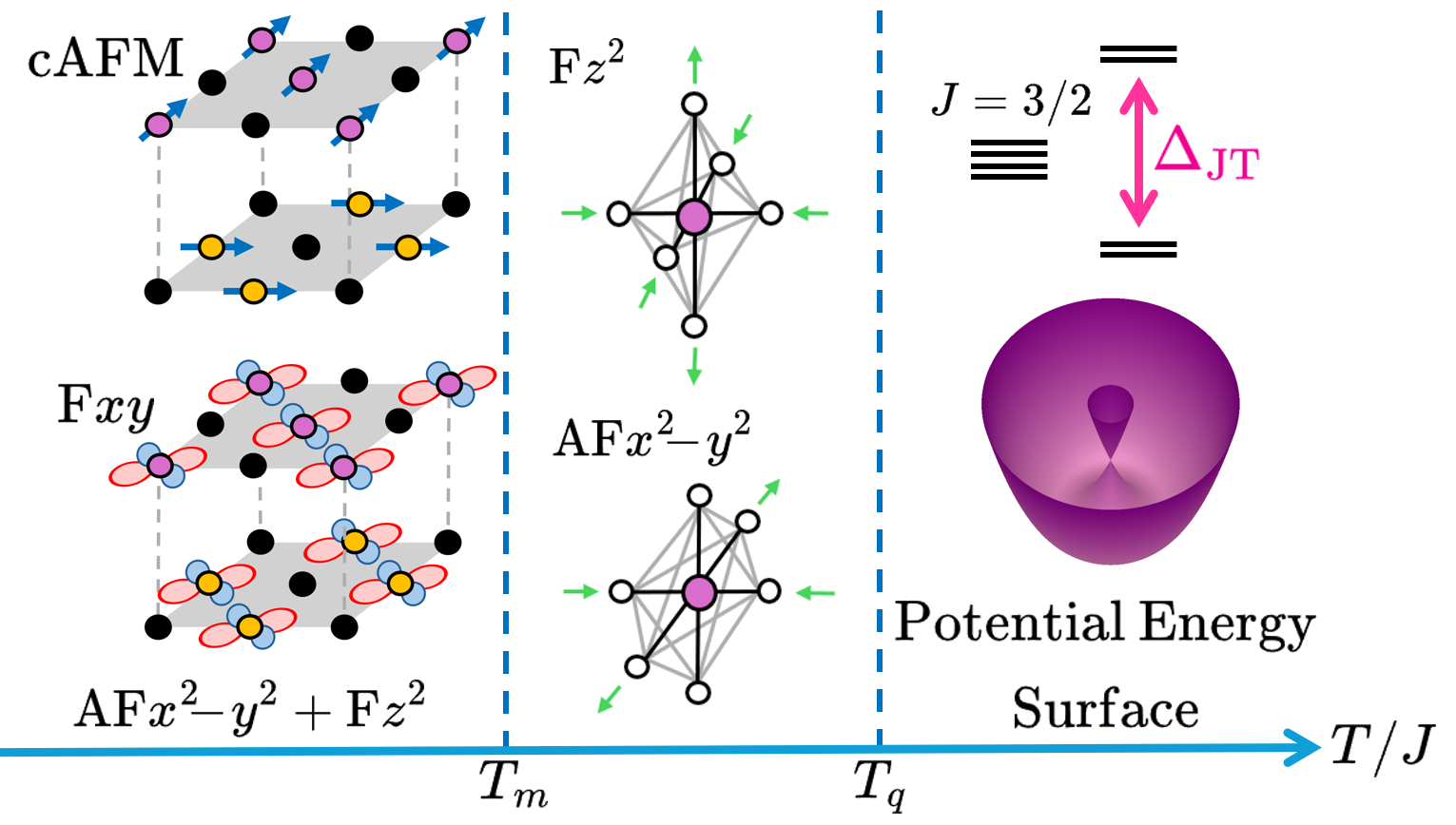}
    \caption{Phases of the $d^1$ double perovskites as a function of temperature, $T/J$. For $T\!<\! T_m$, canted antiferromagnetic (cAFM)
    dipole order coexists with quadrupolar orders: notably antiferro-ordering of $x^2\!\!-\!y^2$ quadrupoles (AF$x^2\!\!-\!y^2$), 
    ferro-ordering of $z^2$ quadrupoles (F$z^2$), and weak ordering of $T_{2g}$ quadrupoles (F$xy$) on the magnetic sites (pink and yellow dots are used to distinguish the different planes). Black dots represent non-magnetic ions. For $T_m \!<\! T \!< \! T_q$, dipolar magnetic order vanishes and pure $E_g$ quadrupolar order persists
    accompanied by depicted lattice distortions $Q_{z^2}$ and $Q_{x^2-y^2}$. For $T \!> \! T_q$, quadrupolar order vanishes but JT splitting of the $j\!=\!3/2$ quartet, with $\Delta_{\rm JT}=\lambda {Q}_{E_g}$, persists to high temperatures; here $\lambda$ is the multipole-phonon
    coupling and ${Q}_{E_g}$ is the lattice distortion. Phonon fluctuations in this
    regime follow a `Rashba'-ring type potential energy surface.}
    \label{fig:schematic}
\end{figure}


In this Letter, we explore the impact of multipole-phonon coupling in
$j=3/2$ Mott insulators on the face-centered cubic (fcc) lattice as
realized in lacunar spinels GaTa$_4$Se$_8$ and GaNb$_4$Se$_8$ and double perovskites
such as Ba$_2$MgReO$_6$ and Ba$_2$NaOsO$_6$.
These materials have been shown to host
multipolar orders \cite{Chen2010Exotic,Hirai2020Detection,Hirai2019Successive,Ishikawa2020Nonmagnetic,Svoboda2021Orbital,Lovesey2021Magnetic,Winkler2022Antipolar,soh2024spectroscopic}, with ample evidence that phonons are crucial for 
selecting the ground state
and shaping the excitation spectrum \cite{Iwahara2018Spin,Liu2018Nature,FMosca2021Interplay,Iwahara2023Vibronic,FMosca2024Interplay,vzivkovic2024dynamic,Iwahara2025Persistent,martinelli2026quadrupole}. Fig.~1 shows a schematic phase diagram, 
highlighting the progression from ordered multipolar ground states, to intermediate
states with coexisting pure quadrupolar orders, to proposed fluctuating JT effects which persist to high temperatures.
Despite these proposed scenarios, several questions remain unexplored. How do thermal JT fluctuations 
impact the local $j=3/2$ spectrum and its temperature evolution? How does the coupling to phonons
impact the thermal entropy of the $j=3/2$ moments, i.e., does the thermally disordered state recover the $k_B \ln 4$
entropy per spin?
How do different experimental measurements which probe the system on different timescales 
sense the impact of the JT phonons as a function of temperature? 
These questions are important to explore in light of the fact that
previous work has primarily focussed on the nature of the
ground states using density functional theory (DFT) \cite{FMosca2021Interplay,FMosca2024Interplay,soh2024spectroscopic,martinelli2026quadrupole} while issues related to thermal fluctuations and coupling to
phonon modes have not been studied beyond mean field theory.
We present a study of these issues
using recently developed $SU(N)$ Monte Carlo algorithms coupled to phonons, and through exact diagonalization
calculations
to uncover the temperature dependent spectral distribution of the $j\!=\!3/2$
multiplet coupled to spatially fluctuating JT distortions.
Our work sheds light on
the interplay of JT phonons and strong SOC in diverse quantum materials.

{\it Model.---} The $j\!=\!3/2$ quartet transforms in the fundamental (defining) 
representation of $SU(4)$. We can thus compactly write the most general exchange Hamiltonian for
$j\!=\!3/2$ Mott insulators using $SU(4)$ generators
\begin{equation}
    H_{\cal T} = \sum_{\br,\br'}\sum_{\mu,\nu} J_{\mu,\nu}(\br,\br') {\cal T}_\mu(\br) {\cal T}_\nu(\br')
\end{equation}
where $J_{\mu,\nu}(\br,\br')$ is the interaction strength between $SU(4)$ generators ${\cal T}_\mu(\br)$ and ${\cal T}_\nu(\br')$.
We emphasize that this does not imply $SU(4)$ symmetry of the Hamiltonian.
Indices $\mu,\nu$ label the $15$ operators which are grouped into 
3 dipoles, 5 quadrupoles, and 7 octupoles (see Supplementary Materials (SM) \cite{suppmat}). Time-reversal symmetry places constraints on bilinear terms in the Hamiltonian: quadrupoles 
do not couple to dipoles or octupoles. 
The intersite exchange couplings $J_{\mu,\nu}(\br,\br')$ may be obtained using tight-binding models
or DFT calculations; for the $d^1$ double perovskites, we retain the
dominant Kugel-Khomskii type spin-orbital superexchange with strength 
$J$ \cite{Chen2010Exotic,Svoboda2021Orbital}, supplemented by quadrupolar couplings inferred
from DFT calculations \cite{FMosca2024Interplay,martinelli2026quadrupole,soh2024spectroscopic,FMosca2021Interplay}.
We adopt the notation $J_{\mu,\nu}(\br,\br') \propto V_{E}$ for quadrupolar interactions of $E_g$ symmetry ($\mu,\nu=z^2,x^2-y^2$) and $J_{\mu,\nu}(\br,\br') \propto V_{T}$ for quadrupolar interactions of $T_{2g}$ symmetry ($\mu,\nu=xy,yz,zx$). 
Details on the form of the exchange Hamiltonian are given in 
the End Matter.

Next, we introduce coupling to JT lattice distortions. There are five relevant lattice
modes $E_g \oplus T_{2g}$ which couple linearly to quadrupoles of matching symmetry. Here, we focus on the two degenerate
$E_g$ modes, noting that they couple more strongly to the quadrupole moments \cite{FMosca2024Interplay}. These 
modes are modelled as Einstein oscillators
\begin{eqnarray}
    H_{\rm ph} = \sum_\br \sum_{\gamma} \left[ \frac{P_\gamma(\br)^2}{2M} + \frac{1}{2} K Q_\gamma(\br)^2 \right]
\end{eqnarray}
where $\gamma=z^2,x^2-y^2$, $P_\gamma(\br)$ and $Q_\gamma(\br)$ are the phonon momentum and coordinate, respectively, at site $\br$, with $M$ being
the mass and $K$ being the stiffness.

Finally, we introduce vibronic quadrupole-phonon couplings $H_{\rm vib}=H_{\rm vib}^{\rm loc} + H_{\rm vib}^{\rm nloc}$, where
symmetry allowed local (loc) and non-local (nloc) couplings, with strengths $\propto \! \lambda$, are
\begin{eqnarray}
\label{eq:sp-ph-local}
    H_{\rm vib}^{\rm loc} \!\!&=&\!\! \lambda \sum_{\langle \br\br' \rangle} \sum_{\gamma} {\cal T}_\gamma(\br) Q_\gamma(\br) \\
    H_{\rm vib}^{\rm nloc} \!\!&=&\!\! g \lambda \!\!\!\!\! \sum_{\langle \br\br'\rangle \in xy} \!\!\!\!\!\!{\mathcal{T}}_{x^2-y^2}(\br) {\cal T}_{x^2-y^2}(\br') (Q_{z^2}(\br) \!+\! Q_{z^2}(\br')) \nonumber \\
    &+&  {\mathrm{symmetry~related~terms~for~{\it yz,zx}~planes}}
\end{eqnarray}
The local vibronic coupling will be stronger than the non-local term
and it will thus have dominant impact on  JT distortions, so we expect $g \! \ll \! 1$; we thus set $g\approx 0.1$.
The non-local vibronic coupling is explicitly written for nearest neighbor sites in the $xy$ plane; neighboring sites in $yz,zx$ planes
can be obtained by $C_3$ rotations about the [111] axis.
This trilinear coupling, not considered in previous work,
leads to intertwining of quadrupolar orders which we explore below using
Monte Carlo simulations.


\begin{figure}[t]
    \centering
    \includegraphics[width=0.98\linewidth]{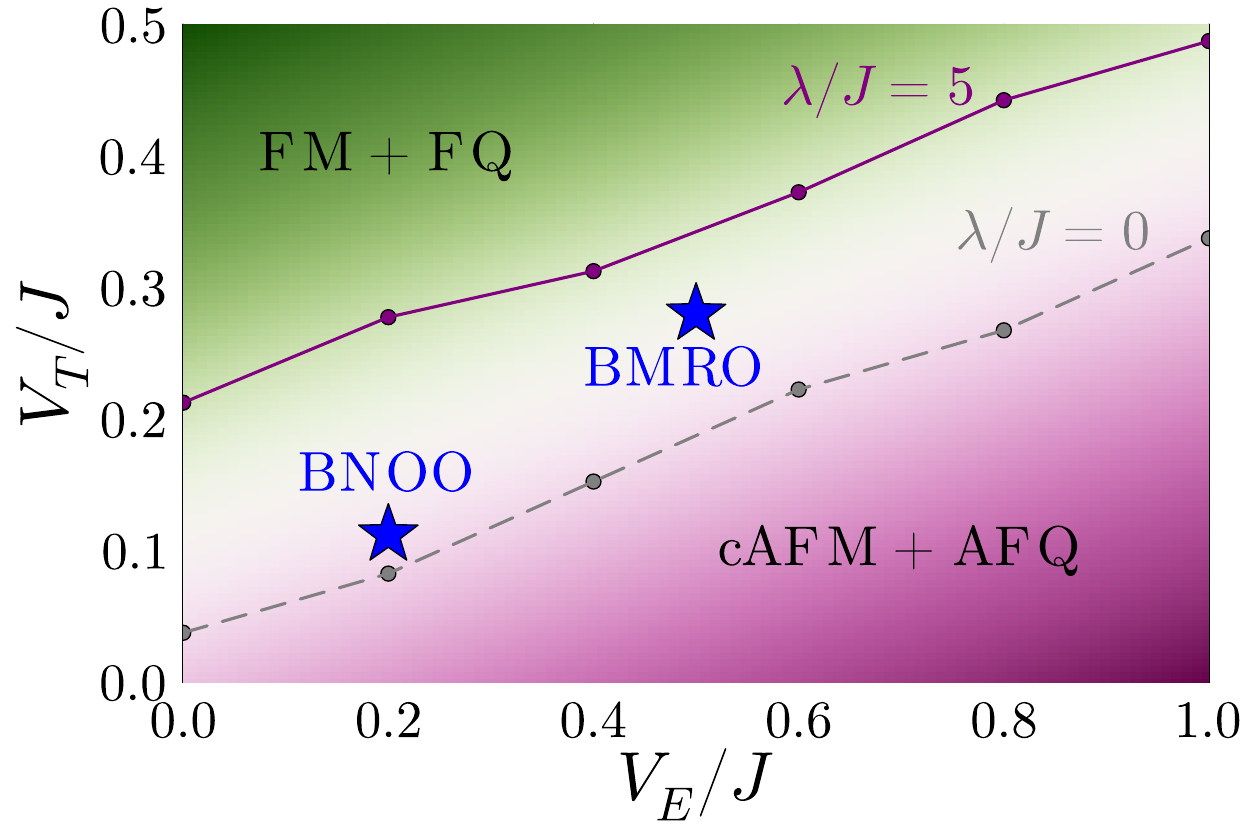}
    \caption{Ground state phase diagram as a function of quadrupole interactions $V_{E}/J$ and $V_{T}/J$. Dashed line shows the phase
    boundary between the FM+FQ phase (green) and cAFM+AFQ phase (pink) in the absence of vibronic coupling $\lambda/J=0$. 
    Incorporating $\lambda/J\! \neq\! 0$ pushes the phase boundary upwards, e.g., solid line shows the displaced boundary for 
    $\lambda/J\!=\! 5.0$. 
    Stars indicate illustrative choices of ($V_{E}/J$, $V_{T}/J$) for {\bmro} and {\bnoo}.
    }
    \label{fig:phaseDiagram}
\end{figure}

{\it Ground state phase diagram.---} Using the $SU(4)$ Monte Carlo method coupled to classical JT phonons, we have
constructed the phase diagram of model $H_{\cal T}$, in the absence of coupling to phonons, as well as in the presence of phonon coupling
described by $H_{\rm{vib}}$  (see End Matter for simulation details).
Fig.~\ref{fig:phaseDiagram} shows the ground state phase diagram for the model as a function of $V_{E}/J$ and $V_{T}/J$. 
At large $V_{T}/J$ we find 
dipolar ferromagnetic (FM) order along the [111] direction with coexisting $T_{2g}$ ferroquadrupolar (FQ) order where
$\langle {\cal T}_{xy}(\br)\rangle=\langle {\cal T}_{yz}(\br)\rangle=\langle {\cal T}_{zx}(\br)\rangle$ at each site. This FM+FQ phase is
consistent with first-principles calculations in the absence of vibronic coupling \cite{FMosca2024Interplay,martinelli2026quadrupole,soh2024spectroscopic,FMosca2021Interplay}. 
Conversely, at large $V_{E}/J$, we find previously proposed
canted antiferromagnetic dipolar order accompanied by dominant antiferro-quadrupolar ${\cal T}_{x^2-y^2}$  order (cAFM+AFQ);
this phase is realized in {\bmro} and {\bnoo} at low temperature \cite{Chen2010Exotic,lu2017magnetism,Hirai2019Successive,Hirai2020Detection,Svoboda2021Orbital,soh2024spectroscopic,martinelli2026quadrupole}.
The dipolar AFM ordering pattern is depicted in Fig.\ref{fig:schematic}, where spins in neighbouring $xy$ planes are each canted in opposite directions away from the [110] axis, yielding a net moment along this axis. Experimental fits estimate canting angles of approximately $40^\circ$ in {\bmro} \cite{Hirai2019Successive} and $67^\circ$ in {\bnoo} \cite{lu2017magnetism}. This cAFM+AFQ phase generically
supports a weak ferroic order of ${\cal T}_{xy}$ quadrupoles.

Turning on the vibronic coupling $\lambda/J \neq 0$ favors the cAFM+AFQ phase at the expense of the FM+FQ phase; this manifests in our calculation as a vertical shift of the phase boundary which moves to higher values of $V_{T}$ with increasing $\lambda$, 
as shown in Fig.~\ref{fig:phaseDiagram} for moderate coupling $\lambda/J=5$, while going to strong vibronic coupling $\lambda/J=15$ pushes the
phase boundary further upwards. This observation that the vibronic coupling modifies the ground state is
consistent with prior first-principles studies for {\bmro} and {\bnoo} \cite{FMosca2021Interplay,FMosca2024Interplay,soh2024spectroscopic,martinelli2026quadrupole}. 
We thus place the parameters for these materials close to the
phase boundary for $\lambda/J=0$, as shown in Fig.~\ref{fig:phaseDiagram}, 
such that coupling to phonons transforms their true ground state from the FM+FQ (for $\lambda/J\!=\! 0$)
to the cAFM+AFQ phase for sufficiently strong $\lambda$.
The position of the phase boundary does not depend strongly on the strength of the {\it non-local} weak
vibronic coupling dictated by $g \ll 1$; importantly, however, this symmetry-allowed trilinear term nucleates
a nonzero ferro-quadrupolar component 
${\cal T}_{z^2}$ in the ground state which is `parasitic' to the primary AFQ ${\cal T}_{x^2-y^2}$ order. Effectively, for
$g \lambda/J \!\neq\! 0$, the AFQ ordering
of ${\cal T}_{x^2-y^2}$ creates an internal uniform field on the ${Q}_{z^2}$ phonon mode,
as seen from the form
of $H^{\rm nloc}_{\rm vib}$,
which in turn creates
a nonzero ${\cal T}_{z^2}$ via $H^{\rm loc}_{\rm vib}$ so the two quadrupolar orders
get intertwined and must coexist. This is consistent with
experiments on {\bmro} \cite{Hirai2020Detection,soh2024spectroscopic}. 

{\it Thermal phase diagram.---} 
Given the coexisting dipolar-quadrupolar orders and lattice distortions in the ground state, it is natural to ask 
how thermal fluctuations impact these orders. Upon heating above $T_m \! \approx\! 18$\,K, 
{\bmro} loses the dipolar magnetic order associated with its cAFM+AFQ ground state, transitioning into an
intermediate phase with pure quadrupolar order composed of staggered ${\cal T}_{x^2-y^2}$ and ferroic ${\cal T}_{z^2}$
as inferred from lattice distortions \cite{Hirai2019Successive,Hirai2020Detection} and measured using REXS \cite{soh2024spectroscopic}. Upon
heating further, these intertwined quadrupolar orders are simultaneously lost beyond $T_q \! \approx \! 33$ K. {\bnoo} has a 
similar phase diagram, but with more closely spaced transitions, $T_m \!\approx\! 7.5$\,K and $T_q \! \approx \! 9.5$\,K \cite{Willa2019Phase}.
While pinning down the precise form
of this quadrupolar order in {\bnoo} has proven difficult,
nuclear magnetic resonance (NMR) experiments on {\bnoo} show definitive evidence of an intermediate purely quadrupolar
phase, dubbed a `broken local point group symmetry' (BLPS) phase \cite{lu2017magnetism}, linked to tetragonal distortions involving
oxygen atom displacements along the bond directions \cite{Liu2018Nature}. Remarkably, recent NMR experiments suggest that the
local JT distortions persist above room temperature \cite{nikolov2026orbitalglassconcealsmissing}.

\begin{figure}
    \centering
    \includegraphics[width=0.98\linewidth]{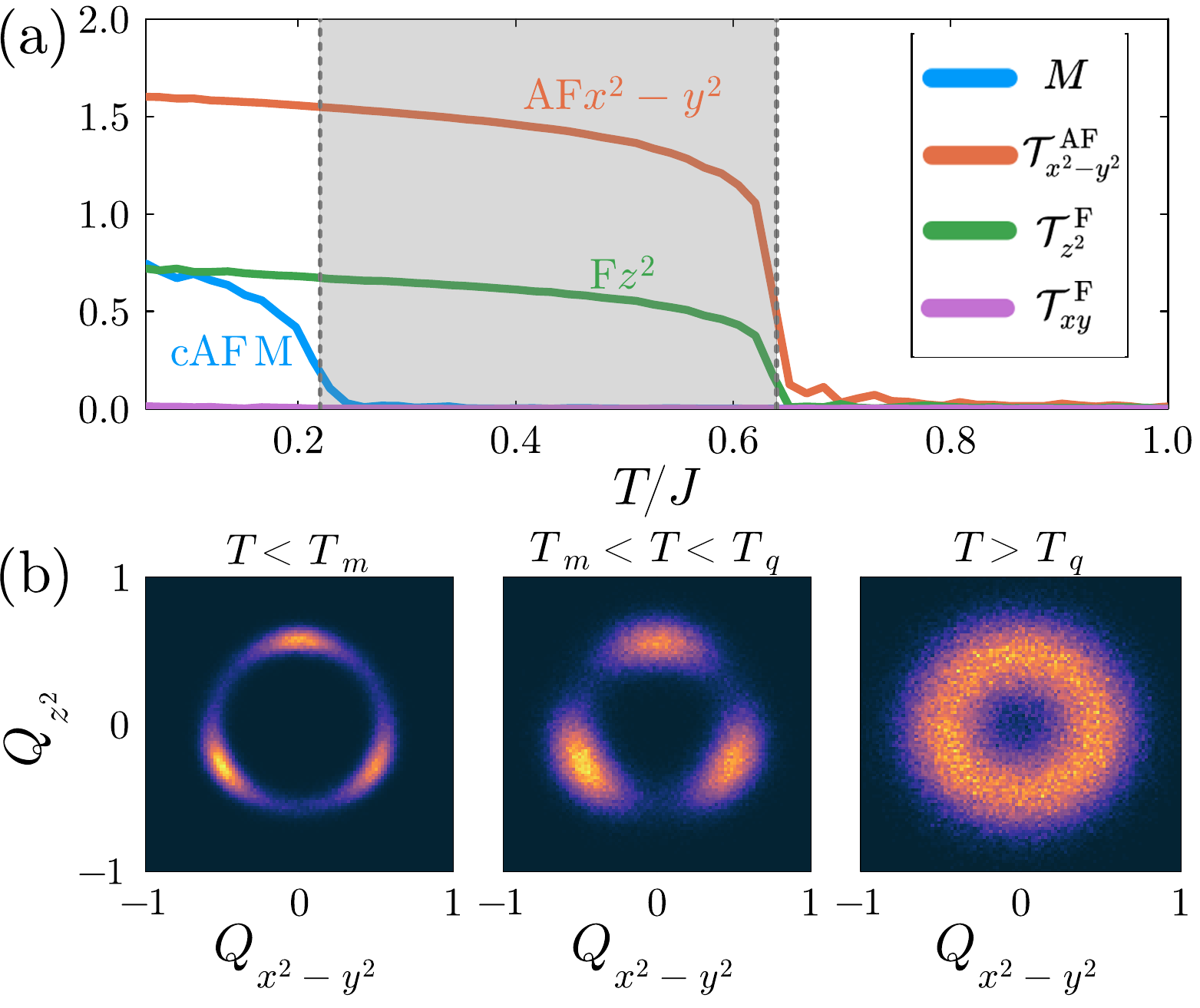}
    \caption{Multipolar orders for $(V_E/J,V_T/J)\!=\! (0.5,0.25)$ and strong vibronic coupling $\lambda/J=15$ relevant to {\bmro},
    and corresponding phonon distributions. (a) Order parameters, showing the high temperature paramagnetic phase,
    an intermediate temperature quadrupolar ordered phase (grey area) $T \!< \! T_q$, and then dipolar magnetic ordering into
    a cAFM [110] order with coexisting quadrupolar orders for $T \!<\! T_m$. (b) Phonon distributions for each phase, sampling over all
    lattice sites and 192 MC configurations. Below $T_q$, the distribution shows three peaks corresponding to tetragonal distortions along three equivalent axes. For $T \!>\! T_q$, the distribution shows a ring structure originating from fluctuations around the `Rashba'-ring potential energy surface
    in Fig. \ref{fig:schematic}.}
    \label{fig:orders}
\end{figure}

We have carried out finite temperature simulations of the multipole-phonon coupled system using the $SU(4)$+phonon approach
for strong vibronic coupling $\lambda/J\!=\! 15$ and fixed $V_T/V_E\!=\! 0.5$.
We discuss results for an illustrative point $(V_E/J,V_T/J)\!=\! (0.5,0.25)$ marked 
as $\bm{\star}$ for {\bmro} in Fig.~\ref{fig:phaseDiagram}; End Matter presents similar results for
$(V_E/J,V_T/J)\!=\! (0.2,0.1)$ for {\bnoo} and for intermediate vibronic coupling $\lambda/J=5$ motivated
by DFT results which argue for weaker phonon-multipole coupling in {\bnoo} \cite{martinelli2026quadrupole}.

Fig.~\ref{fig:orders}(a) shows the various multipolar 
order parameters for $(V_E/J,V_T/J)=(0.5,0.25)$, namely, the net magnetization $M$, the staggered $x^2-y^2$ quadrupole moment ${\cal T}_{x^2-y^2}^{\rm AF}$, and uniform quadrupole moments ${\cal T}_{z^2}^{\rm F}$ and ${\cal T}_{xy}^{\rm F}$ (and equivalent quadrupole moments for equivalent directions of the order).
At low temperature, we find a nonzero
cAFM+AFQ order of the previously discussed ground state, together with an intertwined ferroic ${\cal T}_{z^2}$. 
Upon heating to $T/J \approx 0.25$, the dipolar cAFM order is lost,
but the intertwined staggered ${\cal T}_{x^2-y^2}$ and ferroic ${\cal T}_{z^2}$ persist until $T/J \approx 0.65$, beyond which the
system becomes a paramagnet with no broken symmetries. Setting $J\!=\!5$\,meV, we estimate $T_m\!\approx\! 14$\,K and 
$T_q\!\approx\! 35$\,K, in reasonable agreement with experiments on {\bmro} \cite{Hirai2019Successive,Hirai2020Detection}.
While it has been theoretically
proposed that mixing of the $j\!=\! 3/2$ levels with higher energy spin-orbit split $j\!=\! 1/2$ states, which are separated
by an energy $\zeta_{\rm so} \!\sim\! 0.6$\,eV, could also induce
the coexisting ferroic ${\cal T}_{z^2}$ component \cite{Kubo2023Electronic}, 
the resulting induced ferroic component for $T_m < T < T_q$ appears to be at least
an order of magnitude smaller than the experimental value 
since $J/\zeta_{\rm so} \lesssim 10^{-2}$, 
suggesting that phonons might play a more important role as discussed in our work.

Fig.~\ref{fig:orders}(b) shows the thermal evolution of the corresponding phonon displacements, visualizing the distribution of
$(Q_{x^2-y^2},Q_{z^2})$ across the entire lattice, sampled over $192$ Monte Carlo configurations at each temperature.
At low $T \!<\! T_m$, the phonons
displacements are peaked at three points along the `Rashba'-ring minimum of the potential energy surface
shown in Fig.~\ref{fig:schematic}. These arise from
the staggered ${\cal T}_{x^2-y^2}$, ${\cal T}_{x^2-z^2}$, ${\cal T}_{y^2-z^2}$ antiferro-quadrupolar
orders as we ergodically sample the various allowed broken
symmetry configurations in our finite system size simulations. At intermediate temperatures, $T_m \!<\! T \!<\! T_q$, this
three-peak structure around the `Rashba'-ring continues to exist due to the persistent quadrupolar order although the cAFM order is lost.
Finally, for $T\!>\! T_q$, quadrupolar order is lost and
the system freely fluctuates around the `Rashba'-ring revealing strong thermal
JT fluctuations in the angular direction and weaker fluctuations in the radial 
direction. Despite the symmetric nature of the paramagnetic phase $T\! >\! T_q$, our simulations reveal that local thermally
fluctuating JT distortions persist to high temperature in accord with recently proposed scenarios \cite{vzivkovic2024dynamic}.
We next turn
to implications of this fluctuating JT regime for spectroscopic and thermodynamic probes.

{\it Spectroscopic signatures. ---} 
JT lattice distortions, which break the octahedral point group symmetry of the magnetic site, split the degenerate
$j\!=\!3/2$ quartet into two doublets. 
The splitting $\Delta_{\rm JT}$ has a distribution which reflects 
the distribution of local distortions.
To examine the spectrum of the $j=3/2$ multiplet,
we construct the local Hamiltonian for the multipolar degrees of freedom at each site which arises
from coupling to the local JT phonon distortions. Diagonalizing this local Hamiltonian, we calculate the inhomogeneous
spectral splittings of the $j=3/2$ multiplet and study the temperature-dependent gap distribution.

\begin{figure}
    \centering
    \includegraphics[width=0.98\linewidth]{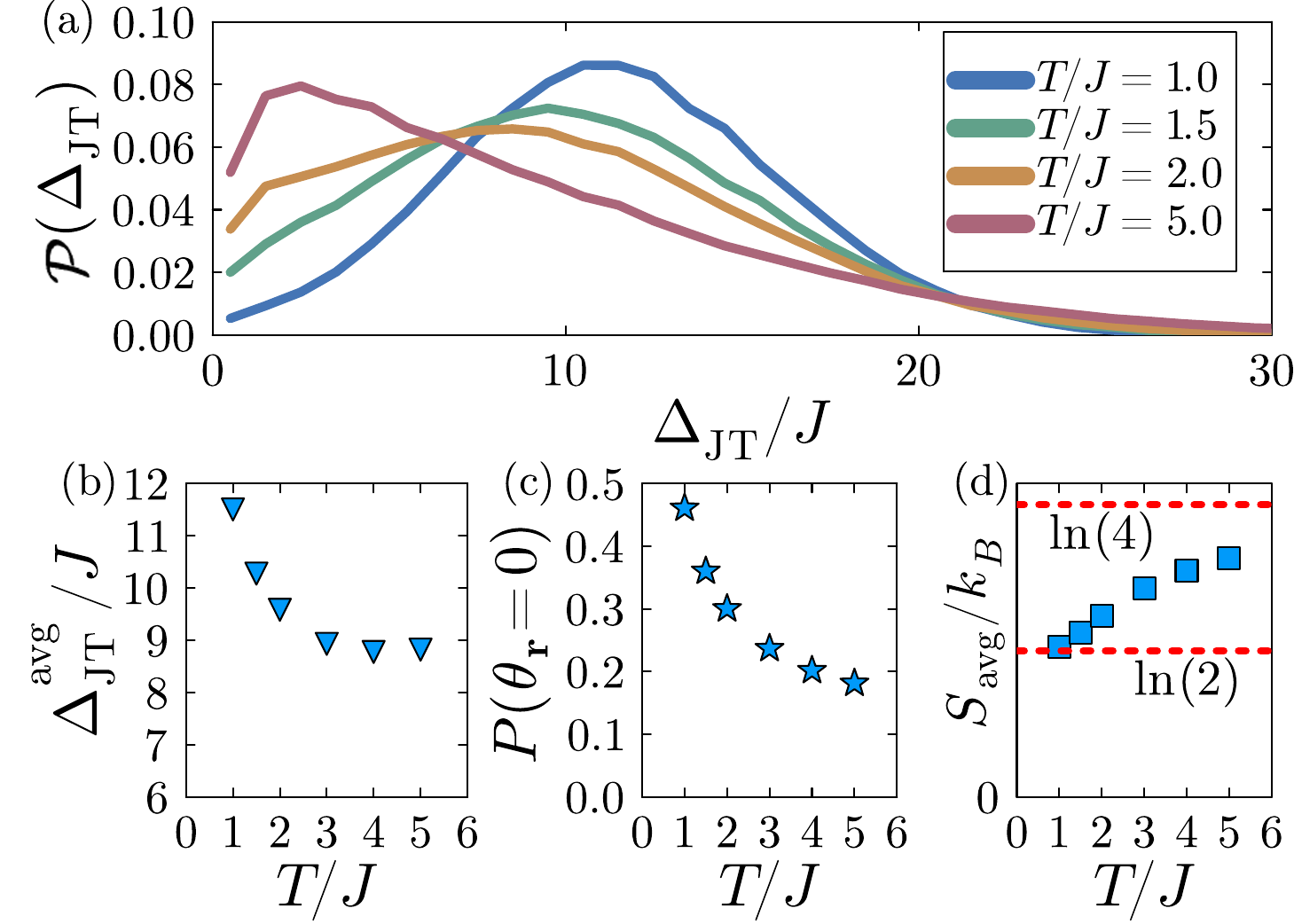}
    \caption{Measures of local JT symmetry breaking at high temperatures for parameters relevant to {\bmro} with strong vibronic coupling $\lambda/J=15$. (a) Distribution of local quartet splitting $\Delta_{\rm JT}$ over a range of temperatures in the paramagnetic regime. The distributions show a maximum at nonzero splitting with a long tail to large $\Delta_{\rm JT}$. (b) Average local quartet splitting as a function of temperature $T/J$. The average splitting is nonzero even for $T \!\gg\! T_q$. (c) Temperature dependence for the height of the distribution of quadrupole-phonon angle, $\cos\theta_\br$ (defined in the text), at $\theta_\br=0$. (d) 
    Entropy $S_{\rm avg}$ of the local $j\!=\!3/2$ quartet averaged over the distribution of splittings in panel (a). 
    }
    \label{fig:entropy}
\end{figure}

We focus here on the puzzling high temperature paramagnetic regime, $T \!>\! T_q$, where we expect the predominant
impact to arise from JT distortions and not intersite multipolar exchange. Fig.~\ref{fig:entropy}(a) 
shows this distribution ${\cal P}(\Delta_{\rm JT})$ as a function of $\Delta_{\rm JT}$ for various temperatures for 
$1 \!<\! T/J \!<\! 5$. In all cases, we find a broad distribution of JT splittings, with the distribution becoming broader
at higher temperatures and the peak of ${\cal P}(\Delta_{\rm JT})$ shifting to lower energies. Fig.~\ref{fig:entropy}(b) shows the average JT splitting
as a function of temperature. Setting $J=5$\,meV as before, the average splitting ranges from $12 J \! \approx \! 60 $\,meV
for $T\!\approx\! 60$\,K to $9 J\!\approx\! 45$\,meV for $T\!\approx\! 300$\,K. For $T \!\approx\! 60$\,K, the broad distribution 
of JT splittings and the average JT splitting is in reasonable agreement
with resonant inelastic x-ray scattering (RIXS) 
experiments \cite{vzivkovic2024dynamic}. However, the RIXS work infers a much weaker temperature dependence of the average
splitting
$\Delta^{\rm avg}_{\rm JT}$ compared with our result in Fig.~\ref{fig:orders}(b) and evidence for vibronic modes; 
we attribute this to our classical
phonon simulations as opposed to quantum phonons
\cite{Iwahara2025Persistent,frontini2024spin,Iwahara2023Excitations}; 
specifically, we expect that the soft angular fluctuations of the phonon mode may be treated 
classically but quantization must be important for the radial amplitude modes.

To see that the spectral gap distribution is tied to the local JT lattice distortions, we calculate a measure of the local quadrupole-phonon locking at individual sites. We define the vector $\vec{Q}_\br=(Q_{x^2-y^2}(\br),Q_{z^2}(\br))$ for JT distortions at a single site $\br$, and a similar vector for $E_g$ quadrupoles $\vec{\cal{T}}_\br=(\langle{\cal T}_{x^2-y^2}(\br)\rangle,\langle{\cal T}_{z^2}(\br)\rangle)$. At each site, we can then quantify the local alignment of the quadrupole and phonon as:
$$\cos\theta_\br=\frac{{\vec{Q}}_\br\cdot{\vec{\cal{T}}}_\br}{{|Q_\br| |\cal{T}}_\br|}.$$
where $\theta_\br$ defines the angle between distortion vector $\vec{Q}(\br)$ and the quadrupole vector ${\cal T}(\br)$.
For $\lambda/J\!=\!15$, the probability distribution $P(\theta_\br)$ turns out to be \emph{strongly} peaked at the angle $\theta_\br=0$, 
indicating robust quadrupole-phonon locking well above $T_q$. 
Fig.~\ref{fig:entropy}(c) shows the temperature dependence of $P(\theta_\br=0)$ for the parameter set for {\bmro} 
from which we observe that a significant $P(\theta_\br=0)$ persists to $T/J \!\sim\! 5$, which appears consistent with experimentally observed splittings and the proposed dynamic Jahn-Teller effect (see SM \cite{suppmat}). 
Thus, while the quadrupoles and phonons may individually appear thermally disordered,
they remain locally correlated; this leads to local symmetry breaking while maintaining {\it average} global cubic symmetry.

{\it The entropy puzzle. ---}
Heat capacity measurements on both {\bmro} and {\bnoo} have attempted to extract the multipole contribution to the temperature 
entropy by subtracting the lattice contribution. Naively, this should lead to an entropy $k_B \ln 4$ per $j\!=\!3/2$ moment for
$T\! \gg\! T_q$ in a cubic system. However, uncertainties in subtracting the phonon contribution in previous works 
appear to yield conflicting results, finding either that the entropy per moment saturates to 
$k_B \ln 2$ \cite{Erickson2007Ferro,vzivkovic2024dynamic} or increasing to a higher value but
still lower than $k_B \ln 4$ \cite{Pasztorova_2023}.


To shed light on the multipolar entropy, we compute the partition function of the local multipolar Hamiltonian discussed 
above with its spatially inhomogeneous distribution of spectral splittings, 
and thus evaluate the $T$-dependent entropy of the multipolar degrees of freedom as shown in Fig.\ref{fig:entropy}(d); 
see SM \cite{suppmat} for details.
We find that the entropy for $T \!>\! T_q$ grows to values larger than $k_B\ln 2$ 
but remains well below $k_B\ln 4$ per site
even out to $T/J\!=\! 5$, in agreement with ref.\cite{Pasztorova_2023}. The fluctuating JT distortions and the resultant spectral
gap distribution thus significantly suppress the
expected $k_B\ln 4$ entropy of the $j\!=\!3/2$ multipolar degrees of freedom.


{\it Discussion.---} We have shown that JT phonons can sculpt the ground state and control thermally fluctuating multipolar orders
in solids with strong SOC, leading to important spectroscopic and thermodynamic signatures. Our work allows us to qualitatively
reconcile RIXS, NMR, and thermodynamic measurements on the double perovskites. RIXS is a fast probe and can thus see spectral 
signatures of the instantaneous phonon distortion
\cite{vzivkovic2024dynamic} which corresponds to a specific point on the `Rashba'-ring 
in Fig.~\ref{fig:orders}(b) and the corresponding JT splitting. While RIXS results have been interpreted in
terms of quantum excitations of the JT mode \cite{Iwahara2025Persistent}, our results show that spatially 
inhomogeneous classical distortions can also play an important role.
In contrast to RIXS, NMR is a much slower probe, which time-averages over the 
angular drift of the quadrupolar order around the `Rashba'-ring. Assuming this angular drift is faster than
the NMR timescales above a certain temperature $T_* \!>\! T_q$, we expect the time-averaged internal quadrupolar fields 
to then vanish at the Na nuclear site for $T \!>\! T_*$, which could be (mis)interpreted as perfect restoration of cubic symmetry
\cite{Liu2018Nature} although recent NMR techniques have revealed local distortions persisting to room temperature \cite{nikolov2026orbitalglassconcealsmissing}.
Finally, thermodynamic measurements have the longest time-scales, so we might naively expect this should agree with NMR results
rather than RIXS. However, unlike NMR,
the specific heat is {\it insensitive} to the `direction' of the
quadrupolar order corresponding to the angular drift
around the `Rashba'-ring for $T > T_q$.
Instead, the entropy suppression
\cite{vzivkovic2024dynamic} below $k_B \ln 4$, is tied to
spectral splittings induced by the nonzero {\it amplitude} of phonon distortions corresponding to
radial localization on the `Rashba'-ring in Fig.~\ref{fig:orders}(b). We stress that a quantum treatment of these
radial modes is important to fully explain the RIXS results and entropy.

While we have focussed here on ordered
double perovskites {\bmro} and {\bnoo} as concrete examples where there is a wealth of recent experimental data, qualitatively
similar results have been found in vacancy ordered halide double perovskites \cite{Pradhan2024Multipolar,Ishikawa2019Ordering,Tehrani2023Charge}. Similarly, the lacunar spinels 
GaTa$_4$Se$_8$ and GaNb$_4$Se$_8$, where Ta or Nb tetrahedra form $j=3/2$ cluster Mott insulators with an fcc lattice structure,
also exhibit a large temperature window above their dimerization transition
with signatures of average cubic symmetry but with evidence of fluctuating local JT distortions \cite{yang2022bond,Yang2024Jahn};
our work is broadly applicable
to this regime although not to their low temperature broken symmetry phases. Extending our work to study the ordered phases of the 
lacunar spinels would be a fruitful next step. Finally, the strong multipole-phonon
coupling explored here may provide the pairing glue for superconductivity in $j=3/2$ multipolar metals \cite{boettcher2016superconducting,brydon2016pairing,roy2019topological,sim2020multipolarsuperconductivity,ishihara2021tuning,Liu2021_KTaO3,villar2021striped,Buessen2021_KTaO3,Liu2023_KTaO3},
which is an important topic for future research.

\acknowledgments
We acknowledge helpful discussions with Claude Ederer, Ivica Zivkovic, and Vesna Mitrovic.
This research was funded by the Natural Sciences and Engineering Research Council of Canada (NSERC) through Discovery grant RGPIN-2026-04578. 
KH acknowledges support from an Ontario Graduate Scholarship (OGS).
RS acknowledges support
from the NSERC through a Canada Graduate Research
Scholarship (CGRS-D).
Computations were performed on the Trillium supercomputer at the SciNet HPC Consortium. SciNet is funded by Innovation, Science and Economic Development Canada; the Digital Research Alliance of Canada; the Ontario Research Fund: Research Excellence; and the University of Toronto.

\bibliography{main,main_2}

\clearpage

\section*{End Matter}

\subsection{Multipolar exchange Hamiltonian}

Early theoretical work on the $j=3/2$ systems modelled the spin-orbital ground state using superexchange derived from interactions with oxygen orbitals, both in the finite \cite{Romhanyi2017Spin,Svoboda2021Orbital} and infinite \cite{Chen2010Exotic} SOC limits. 
This takes the form of a a spin-orbital superexchange (SE) Hamiltonian of the Kugel-Khomskii type
\begin{eqnarray}
    H_{SE}\!\!&=&\!\!\! \sum_\alpha\sum_{\nnSum}\Big[J_1(3/4\!+\!\SSdot)(n_\alpha(\br)\!-\!n_\alpha(\br'))^2 \nonumber
    \\\!&+&\!\!\! J_2(1/4-\SSdot) (n_\alpha(\br)\!+\!n_\alpha(\br'))^2 \nonumber
    \\
    \!&+&\!\!\!J_3(1/4-\SSdot)(n_\alpha(\br) n_\alpha(\br')) \Big]
\end{eqnarray}
where $J_1 = -J/4(1-3\eta)$, $J_2 = -J/4(1-\eta)$, $J_3 = J\eta/((1-\eta)(1+2\eta))$, $J=4t^2/U$, $\eta = J_H/U$, $\bf S(\br)$ is the spin on site $\br$ and $n_\alpha(\br)$ is the orbital occupation on site $\br$ of symmetry $\alpha=xy,yz,zx$. Notably, these models also consider an electric quadrupolar interaction rooted in orbital-orbital repulsion given by
\begin{eqnarray}
    H_{E}=&V_E&\sum_\alpha\sum_{\nnSum}\Big[\frac{9}{4}n_\alpha(\br) n_\alpha(\br') \nonumber
    \\
    \!&-&\! \frac{4}{3}(n_\beta(\br)\!-\!n_\gamma(\br))(n_\beta(\br')\!-\!n_\gamma(\br')) \Big]
\end{eqnarray}
where $V_{E}$ is the strength of the interaction (referred to as simply $V$ in refs. \cite{Chen2010Exotic} and \cite{Svoboda2021Orbital}). 
In order to express these interactions in terms of the generators of the $SU(4)$ representation, we project the spin and orbital operators to the $j=3/2$ manifold and use the $SU(4)$ generators ${\cal T}_\gamma$, the details of which are given in the SM \cite{suppmat}. In this way, $H_E$ can in general be expressed in the $xy$ plane as 
\begin{eqnarray}
    H_E (xy) = V_E \sum_{\langle \br \br'\rangle \in xy}\Big[-\frac{1}{6}({\cal T}_{z^2}(\br)+{\cal T}_{z^2}(\br')) \nonumber
    \\
    +\frac{1}{9}{\cal T}_{z^2}(\br){\cal T}_{z^2}(\br')-\frac{16}{81}{\cal T}_{x^2-y^2}(\br){\cal T}_{x^2-y^2}(\br')\Big]
\end{eqnarray}
where terms in other planes $H_E(xz), H_E(yz)$
are related via symmetry and the coefficients are determined from the transformation given in the SM \cite{suppmat}.

This theoretical model has been largely successful in reproducing a number of key experimental observations in the $d^1$ double perovskites, including both the coexisting magnetic and quadrupolar orders and the intermediate quadrupolar order only regime. However, the scope is limited to the $E_g$ quadrupoles in isolation without the symmetry-permitted coupling lattice distortions, or $T_{2g}$ inter-site quadrupolar interactions. Recent first-principles calculations have computed the full intersite exchange interaction (IEI) matrix, which reveals an additional ferromagnetic interaction between $T_{2g}$ quadrupoles \cite{FMosca2024Interplay,martinelli2026quadrupole,soh2024spectroscopic,FMosca2021Interplay}. 
In order to generalize the model above ($H_{\rm SE}+H_E$), we introduce another quadrupolar interaction inspired by first principles calculations, which is ferromagnetic between $T_{2g}$ quadrupoles. In the $xy$ plane,

\begin{eqnarray}
    H_T(xy) &=& -V_T\sum_{\langle \br \br'\rangle \in xy}\Big[\delta {\cal T}_{xy}(\br) {\cal T}_{xy}(\br') \nonumber
    \\
    &+&{\cal T}_{yz}(\br) {\cal T}_{yz}(\br')+{\cal T}_{xz} (\br){\cal T}_{xz}(\br') \Big]
\end{eqnarray}
with corresponding symmetry related terms $H_T(xz), H_T(yz)$ for the $yz,xz$ planes, where ${\cal T}_\alpha(\br)$ is the quadrupolar operator projected to the $j=3/2$ manifold, and $\delta=0.77$ following from ref. \cite{FMosca2024Interplay}. 
Thus, $H_{\cal T} = H_{SE} + H_{E} + H_T$ defines our multipolar Hamiltonian, which is qualitatively similar to the intersite exchange interaction $H_{\rm IEI}$ determined from first principles \cite{FMosca2024Interplay,martinelli2026quadrupole,soh2024spectroscopic,FMosca2021Interplay}. We report all parameters in units of the superexchange interaction $J$, which has previously been crudely estimated as 
$\sim \! 3$ meV \cite{Svoboda2021Orbital}.

\subsection{Details of $SU(N)$ Monte Carlo simulations}
The numerical results presented in Fig.~\ref{fig:phaseDiagram}, Fig.~\ref{fig:orders} and Fig.~\ref{fig:entropy} are obtained 
using the $SU(N)$+phonon Monte Carlo (MC) simulation technique we have developed for studying coupled multipole-phonon models
\cite{hart2024phonon,sutcliffe2025n}.

For the $j=3/2$ sector, the local multipolar degrees of freedom are represented by $SU(4)$ coherent states
\begin{eqnarray}    
\ket{\psi_\br}= \sum_{m} a_m(\br) |m\rangle
\end{eqnarray} 
where $m=(-3/2,-1/2,1/2,3/2)$, and $a_m(\br)$ are complex numbers constrained by overall
normalization $\sum_m |a_m(\br)|^2=1$ and the irrelevance of phase. 
The mean field variational wavefunction  can be written as $|\Psi\rangle=\otimes_\br |\psi_\br\rangle$, while
the $E_g$ phonons are treated classically with coordinates $Q_\alpha(\br)$ and momentum $P_\alpha(\br)$ where $\alpha$ labels the two degenerate modes. For classical phonons, we can integrate out the momenta independently in the partition function, so we only need to Monte Carlo sample the phonon coordinates $Q_\alpha(\br) \equiv (Q_{z^2}(\br),Q_{x^2-y^2}(\br))$.

The configurations 
$(\{a_m(\br)\},\{Q_\alpha(\br)\})$ are sampled so probabilities
${\cal P} \propto \exp[-E/k_B T]$, where
\begin{eqnarray}
    E \equiv E[\{a_m(\br)\},\{Q_\alpha(\br)\}] = \langle \Psi,Q | H | \Psi,Q\rangle.
\end{eqnarray} 
The Metropolis MC updates for complex coefficients $\{a_m(\br)\}$ is done by selecting new complex numbers which preserve normalization. The phonons updates are done by choosing new
coordinates $Q_\alpha(\br) \in [-Q_{\rm max},Q_{\rm max}]$ where the cutoff $Q_{\rm max} = 10$ is chosen to be much larger than the typical
distortions. In each sweep, we update all multipolar degrees of freedom followed by updating all phonon degrees of freedom.
The simulations are done for system sizes of up to 1000 sites, with one $j=3/2$ degree of freedom and
two phonon modes per site. We ran the simulations for the ground state phase diagram (Fig. \ref{fig:phaseDiagram}) and temperature dependent magnetic and quadrupolar orders (ie. Fig. \ref{fig:orders}(a)) for $4\times10^5$ sweeps with parallel tempering. The simulations for the phonon distributions (ie. Fig. \ref{fig:orders}(b)) were averaged over 192 configurations each run for $1\times10^5$ sweeps at a fixed temperature.

\subsection{Results for parameter values for {\bnoo}}
\begin{figure}[b]
    \centering
    \includegraphics[width=0.8\linewidth]{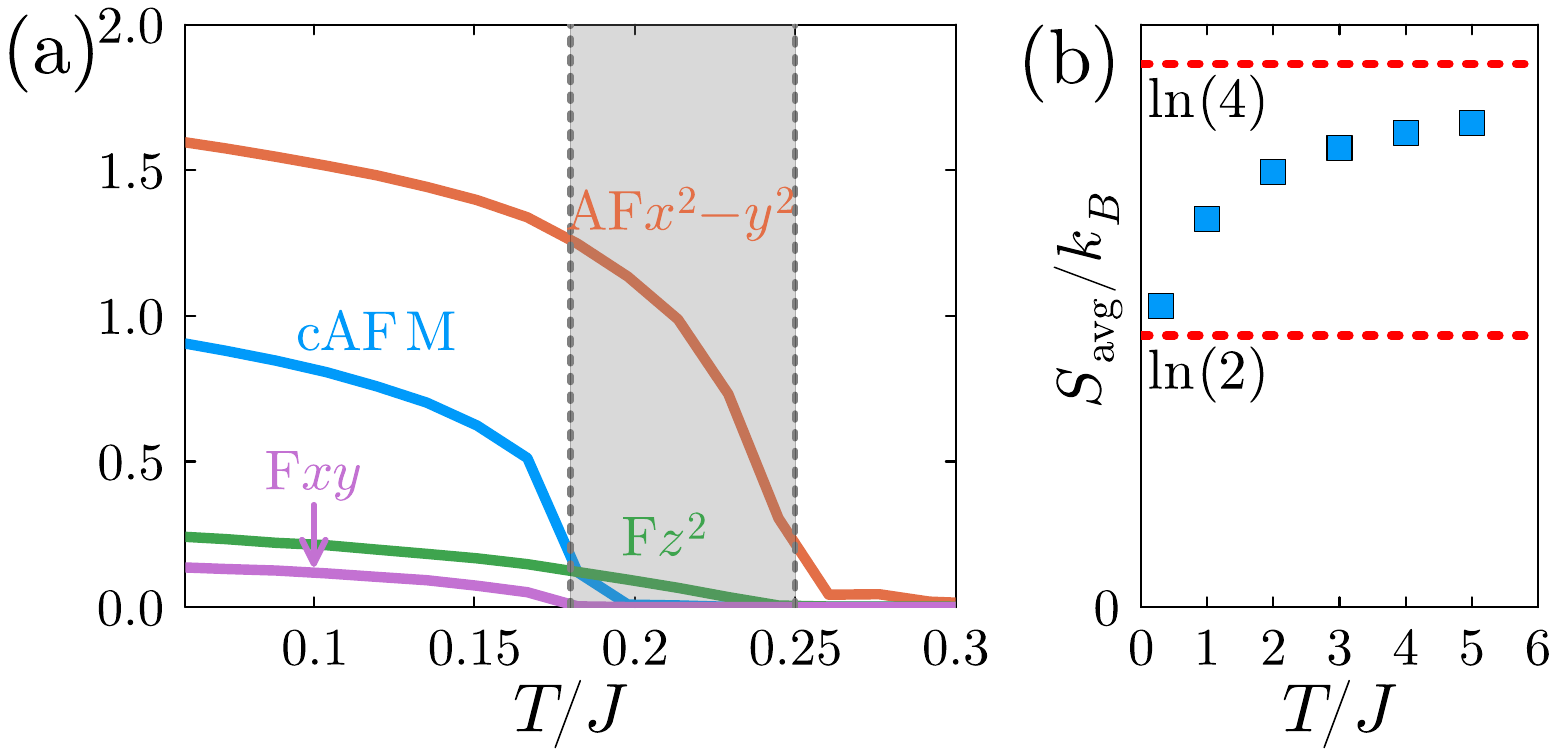}
    \caption{(a) Finite temperature orders with weak vibronic coupling $\lambda/J = 5$ for \bnoo \, ($V_E/J=0.2,V_T/J=0.1$). The low $T$ phase exhibits coexisting cAFM, staggered $x^2-y^2$, ferroic $z^2$  and ferroic $xy$ orders. Above $T_m$, the quadrupolar order persists and vanishes above $T_q$. (b) Average entropy $S_{\rm avg}$ as a function of $T/J$ in the paramagnetic phase.}
    \label{fig:BNOO}
\end{figure}

We present similar results to those presented in Fig. \ref{fig:orders}, for {\bnoo} parameters ($V_E/J\!=\!0.2,V_T/J\!=\!0.1$) in Fig. \ref{fig:BNOO}. It has been previously reported that {\bnoo} exhibits a weaker vibronic coupling compared to {\bmro} \cite{martinelli2026quadrupole}, thus we use a vibronic coupling of $\lambda/J=5$.
The smaller value of $V_E$ and $\lambda$ lead to a lower quadrupolar transition temperature $T_q$, while $T_m$ does not change significantly. Setting $J\!=\!3$ meV, these results lead to transition temperatures $T_m \!\approx\! 0.18J\! \approx \!6$\,K, and $T_q \!\approx\! 0.25 \!\approx\! 9$\,K, in good agreement with reported transition temperatures for \bnoo \cite{Erickson2007Ferro,Willa2019Phase}. Panel (b)
shows the high-$T$ entropy for these parameters, showing a faster approach to $k_B\ln4$ while still remaining below the expected entropy for the quartet at $T/J \!=\! 5 \!\gg\! T_q$.

\subsection{Intermediate quadrupole-phonon coupling}

\begin{figure}[t]
    \centering
    \includegraphics[width=0.85\linewidth]{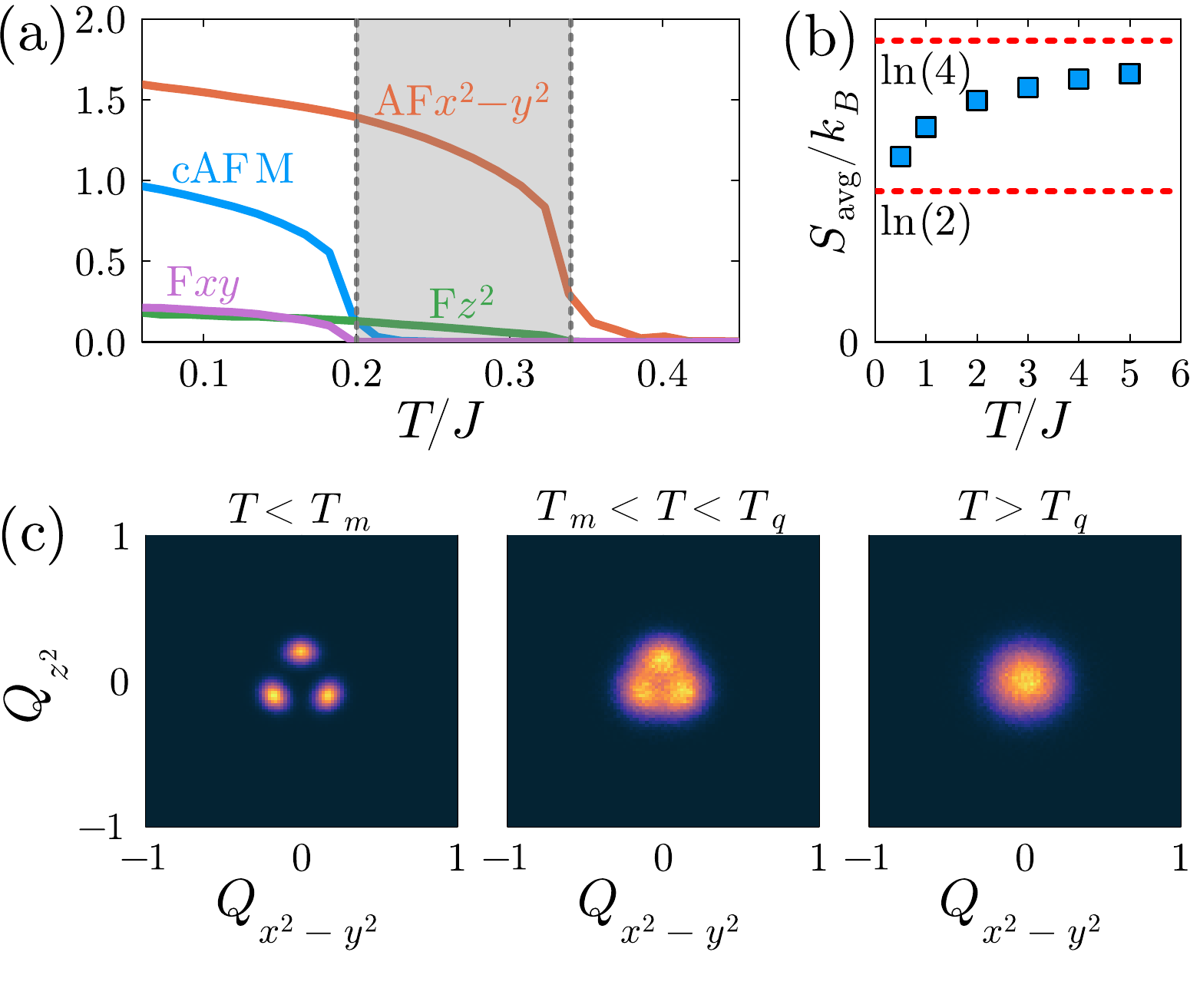}
    \caption{(a) Nonzero $T$ orders with vibronic coupling $\lambda/J = 5$. Similar to the strong coupling case, the low $T$ phase exhibits coexisting cAFM, staggered $x^2-y^2$, ferroic $z^2$  and ferroic $xy$ orders. Above $T_m$, the quadrupolar order persists and vanishes above $T_q$. (b) Average entropy $S_{\rm avg}$ as a function of $T/J$ in the paramagnetic phase. (c) Corresponding phonon distributions for the magnetically ordered phase ($T<T_m$), the intermediate quadrupolar ordered phase ($T_m<T<T_q$) and the paramagnetic phase ($T>T_q$).}
    \label{fig:ordersLargeL}
\end{figure}

To shed light on the role of the vibronic coupling on the thermal phase diagram, the magnetic and quadrupolar orders with weaker
$\lambda/J=5$, along with the corresponding phonon distributions, are presented in Fig. \ref{fig:ordersLargeL}.
We note that the phase progression is very similar to that of the case with $\lambda/J=15$, with some notable differences. First, while the magnetic transition temperature $T_m$ does not significantly change, the quadrupolar transition temperature $T_q$ decreases due to the fact that the vibronic coupling  helps to further stabilize the $E_g$ quadrupolar order. 
Along with the cAFM order, a weak ferroic ordering of $T_{2g}$ quadrupoles, ${\cal T}_{xy}^{F}$ (and equivalent $T_{2g}$ quadrupoles), also develops. In contrast to the $T_{2g}$ quadrupolar order accompanying the FM+FQ phase, this ordering involves only a single $T_{2g}$ quadrupole of representation corresponding to the plane of the magnetic order (ie. ${\cal T}_{xy}^F \neq 0$ for magnetic order in the $xy$ plane). This small $T_{2g}$ order is present even when $V_{T}/J = 0$, indicating that it is not an order stabilized by quadrupolar interactions, but rather arises as a consequence of the cAFM order. Correspondingly, we expect enhanced $T_{2g}$ quadrupolar fluctuations above $T_m$.
While there are quantitative differences for $\lambda/J=5$, these results still match qualitatively with previous experimental and theoretical results on {\bmro} and {\bnoo}.

The phonon distributions show three peaks below $T_m$ that correspond to equivalent tetragonal distortion directions, as is the case for $\lambda/J=15$ (Fig. 3 of the main text). These peaks broaden as temperature is increased above $T_q$,  and eventually they merge into a single peak centred at $\vec{Q}_\br=(0,0)$. The key difference between this result and that of the main text is the lack of a ring structure for $\lambda/J=5$. This is a direct consequence of the `Rashba' potential energy surface, the depth of which is determined by the coupling $\lambda$. Since the weaker vibronic coupling has a shallower potential, the thermal fluctuations above $T_q$ are strong enough to overcome the barrier at the centre (the conical intersection in Fig. 1 of the main text) and consequently results in a distributions that is peaked at $\vec{Q}_\br=(0,0)$. In panel (b), we show the high-temperature entropy for these parameters and find that the entropy more closely approaches the expected $k_B\ln4$ entropy of a quartet but similarly to Fig.\ref{fig:entropy}.(d), remains below the expected value.

\end{document}


\preprint{APS/123-QED}
\title{Supplementary Material: \\Jahn-Teller effect in $j\!=\!3/2$ Mott insulators: Ground states and thermal fluctuations}
\author{Kahleen Hart}
\thanks{These authors contributed equally to this work.}
\affiliation{Department of Physics, University of Toronto, 60 St. George Street, Toronto, ON, M5S 1A7 Canada}
\author{Ruairidh Sutcliffe}
\thanks{These authors contributed equally to this work.}
\affiliation{Department of Physics, University of Toronto, 60 St. George Street, Toronto, ON, M5S 1A7 Canada}
\author{Arun Paramekanti}
\email{arun.paramekanti@utoronto.ca}
\affiliation{Department of Physics, University of Toronto, 60 St. George Street, Toronto, ON, M5S 1A7 Canada}

\date{\today}

\maketitle

\beginsupplement

\section{$SU(4)$ Generators and $j=3/2$ Multipolar Operators}

As discussed in the main text, the model Hamiltonian is projected to the $j=3/2$ manifold using the $SU(4)$ generators $\cal{T}_\gamma$. This projection gives the following transformations \cite{Chen2010Exotic} for the spin and orbital operators:
\begin{equation}
    S_x = \frac{1}{3}{\cal T}_x; \quad S_y = \frac{1}{3}{\cal T}_y; \quad S_z = \frac{1}{3}{\cal T}_z
\end{equation}
\begin{eqnarray}
    n_{xy} = \frac{1}{3}-\frac{2}{9}{\cal T}_{z^2};\quad 
    n_{yz} = \frac{1}{3}+\frac{1}{9}{\cal T}_{z^2}-\frac{1}{3\sqrt{3}}{\cal T}_{x^2-y^2};\quad 
    n_{zx} = \frac{1}{3}+\frac{1}{9}{\cal T}_{z^2}+\frac{1}{3\sqrt{3}}{\cal T}_{x^2-y^2} \nonumber
    \\
    S_xn_{yz} = \frac{1}{15}{\cal T}_x-\frac{2}{15}{\cal T}_{x^3};\quad 
    S_xn_{zx} = \frac{2}{15}{\cal T}_x+\frac{1}{15}{\cal T}_{x^3}-\frac{1}{3\sqrt{15}}{\cal T}_{xy^2+z^2x};\quad 
    S_xn_{xy} = \frac{2}{15}{\cal T}_x+\frac{1}{15}{\cal T}_{x^3}+\frac{1}{3\sqrt{15}}{\cal T}_{xy^2+z^2x} \nonumber
    \\
    S_yn_{yz} = \frac{2}{15}{\cal T}_y+\frac{1}{15}{\cal T}_{y^3}+\frac{1}{3\sqrt{15}}{\cal T}_{yz^2+x^2y};\quad 
    S_yn_{xz} = \frac{1}{15}{\cal T}_y-\frac{2}{15}{\cal T}_{y^3};\quad 
    S_yn_{xy} = \frac{2}{15}{\cal T}_y+\frac{1}{15}{\cal T}_{y^3}-\frac{1}{3\sqrt{15}}{\cal T}_{yz^2+x^2y} \nonumber
    \\
    S_zn_{yz} = \frac{2}{15}{\cal T}_z+\frac{1}{15}{\cal T}_{z^3}-\frac{1}{3\sqrt{15}}{\cal T}_{zx^2+y^2z};\quad 
    S_zn_{xz} = \frac{2}{15}{\cal T}_z+\frac{1}{15}{\cal T}_{z^3}+\frac{1}{3\sqrt{15}}{\cal T}_{zx^2+y^2z};\quad 
    S_zn_{xy} = \frac{1}{15}{\cal T}_z-\frac{2}{15}{\cal T}_{z^3}
\end{eqnarray}
The $SU(4)$ generators ${\cal T}_\gamma$ in terms of the $J=3/2$ operators is given in Table \ref{table:operators}. After projecting to the $J=3/2$ basis, one of the octupole operators, ${\cal T}_{xyz}$, does not enter the exchange interactions, though it is still an allowed multipole within the $J=3/2$ manifold.

\begin{table}[h!]
\centering
\begin{tabular}{|c | c | c|} 
 \hline
 Multipole moment & SU(4) generators & $J=3/2$ operators \\ [0.5ex] 
 \hline\hline
 Dipole & ${\cal T}_x$ & $J_x$  \\  [1ex] 
 & ${\cal T}_y$ & $J_y$  \\  [1ex] 
 & ${\cal T}_z$ & $J_z$ \\  [1ex] 
 \hline
 Quadrupole & ${\cal T}_{x^2-y^2}$ & $J_x^2 - J_y^2$  \\ [1ex] 
 & ${\cal T}_{z^2}$ & $\frac{1}{\sqrt{3}}(2J_z^2-J_x^2-J_y^2)$  \\ [1ex] 
 & ${\cal T}_{xy}$ & $J_xJ_y + J_yJ_x$  \\ [1ex] 
 & ${\cal T}_{yz}$ & $J_yJ_z + J_zJ_y$  \\ [1ex] 
 & ${\cal T}_{zx}$ & $J_zJ_x + J_xJ_z$  \\ [1ex] 
 \hline 
 Octupole & ${\cal T}_{xyz}$ & $\frac{\sqrt{15}}{6}\overline{J_xJ_yJ_z}$ \\ [1ex] 
 & ${\cal T}_{x^3}$ & $J_x^3 - \frac{1}{2}(\overline{J_xJ_y^2} + \overline{J_z^2J_x})$  \\  [1ex] 
 & ${\cal T}_{y^3}$ & $J_y^3 - \frac{1}{2}(\overline{J_yJ_z^2} + \overline{J_x^2J_y})$  \\ [1ex] 
 & ${\cal T}_{z^3}$ & $J_z^3 - \frac{1}{2}(\overline{J_zJ_x^2} + \overline{J_y^2J_z})$  \\ [1ex] 
 & ${\cal T}_{xy^2+z^2x}$ & $\frac{\sqrt{15}}{6}(\overline{J_xJ_y^2} + \overline{J_z^2J_x})$  \\ [1ex] 
 & ${\cal T}_{yz^2+x^2y}$ & $\frac{\sqrt{15}}{6}(\overline{J_yJ_z^2} + \overline{J_x^2J_y})$  \\ [1ex] 
 & ${\cal T}_{zx^2+y^2z}$ & $\frac{\sqrt{15}}{6}(\overline{J_zJ_x^2} + \overline{J_y^2J_z})$  \\ [1ex] 
 \hline
\end{tabular}
\caption{Table of equivalence between $SU(4)$ generators and $J=3/2$ operators.}
\label{table:operators}
\end{table}

\section{Distribution of Quadrupole-Phonon Locking}

\begin{figure}[h]
    \centering
    \includegraphics[width=0.65\linewidth]{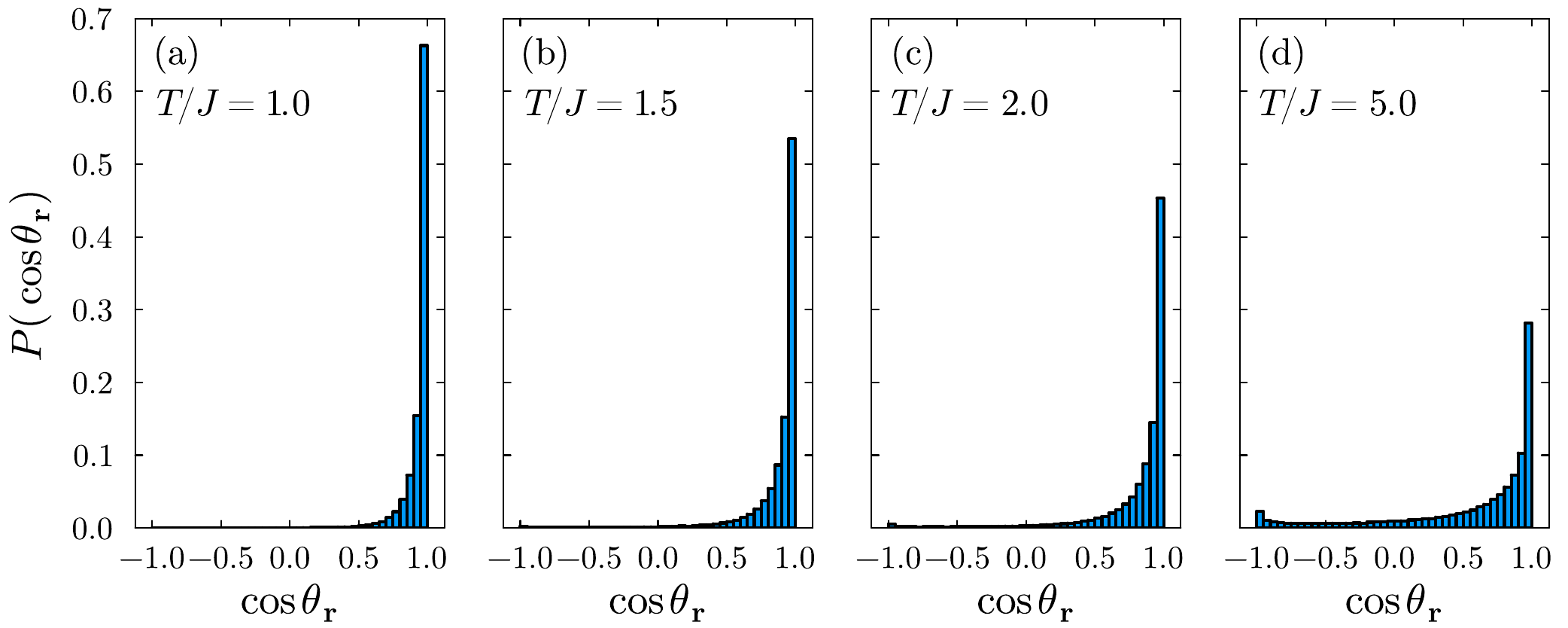}
    \caption{Distribution of the phonon-quadrupole locking, $\cos (\theta_\br)$, as defined in the text, for {\bmro} parameters with $\lambda/J=15$. Distributions for a range of temperatures with $T>T_q$ are shown, each with 192 Monte Carlo configurations. Note that even at the highest temperature shown: $T/J=5.0 \gg T_q$, the distribution remains strongly peaked at $\theta_\br=0$.  }
    \label{fig:phononLocking}
\end{figure}

In Fig. \ref{fig:phononLocking}, we show the distribution of quadrupole-phonon locking as a function of the angle $\cos \theta_\br$ (defined in the main text). We observe a strong peak at $\cos \theta_\br = 1$, where the quadrupole vector $\vec{\cal T}_\br$ and phonon vector $\vec{Q}_\br$ are aligned locally. As temperature is increased, the height of this peak decreases, as shown in Fig. 4 of the main text, and the distribution gains a longer tail away from $\theta_\br=0$. We treat this locking angle $\theta_\br$ as a measure of the correlation between the quadrupole and phonon at each site, indicative of the dynamic Jahn-Teller effect and a direct result of the linear vibronic coupling in the system.

\section{High temperature Entropy}

We consider the system in the high temperature regime ($T>T_q$), where there is no magnetic or quadrupolar order, and as such we can treat each site as independent. Each site hosts a local $j=3/2$ quartet which is split into two doublets by the vibronic coupling $H_{\rm vib}$. If we assume that the lower doublet has an energy $\varepsilon$ and the higher doublet has an energy $\varepsilon+\Delta$, then the partition function for the local system is
$$
Z = \sum_{E_i} e^{-\beta E_i} = 2e^{-\beta \varepsilon}(1+e^{-\beta \Delta})
$$
where $\beta = 1/k_B T$. We can then write down the probability of he system being in a given state as
$
p_i = e^{-\beta E_i}/Z
$
and we can thus compute the entropy as $S=-k_B \sum_i p_i \ln p_i$. since each level is a doublet, we can write the entropy simply as
$
S=-k_B (2p_l \ln p_l + 2p_h \ln p_h)
$
where $p_l$ and $p_h$ are the probabilities of being in the lower and higher doublets, respectively. Thus our entropy is given by
\begin{eqnarray}
    S = -k_B\Big(&2&\left[\frac{1}{2(1+e^{-\beta \Delta})} \ln \frac{1}{2(1+e^{-\beta \Delta})}\right] 
    \\
    + &2&\left[\frac{e^{-\beta\Delta}}{2(1+e^{-\beta \Delta})} \ln \frac{e^{-\beta\Delta}}{2(1+e^{-\beta \Delta})}\right] \Big) \nonumber
    \\
    = k_B\ln &2& + k_B\ln (1+e^{-\beta \Delta}) + \frac{1}{(e^{\beta \Delta}+1)} \frac{\Delta}{T}
\end{eqnarray}
Note that the entropy does not depend on the lower energy level $\varepsilon$ and only on the gap between the levels $\Delta$. In the limit $T \rightarrow \infty$, the entropy is $S = k_B\ln 4$. Conversely, in the limit $T \rightarrow 0$, the entropy is $S=k_B \ln 2$. In order to compute the entropy for the entire system, we take this local entropy and average it over the lattice using the distribution of splittings.

\bibliography{main, main_2}

%% file: defns.tex
\newcommand{\be}{\begin{equation}}
\newcommand{\ee}{\end{equation}}

\newcommand{\br}{{\bf{r}}}

\newcommand{\beal}{\begin{align}}
\newcommand{\eeal}{\end{align}}

\renewcommand{\vec}[1]{\mathbf{#1}}

\def\bmro{Ba$_2$MgReO$_6$}
\def\bnoo{Ba$_2$NaOsO$_6$}

\newcommand{\nnSum}{\langle \br \br' \rangle\in \alpha}
\newcommand{\SSdot}{{\bf {S}(\br)\cdot\bf {S}(\br')}}